\documentclass[preprints,review,accept,moreauthors,pdftex]{Definitions/mdpi} 

\firstpage{1} 
\pubvolume{xx}
\issuenum{1}
\articlenumber{5}
\pubyear{2019}
\copyrightyear{2019}
\history{Received: date; Accepted: date; Published: date}

\Title{GHz Superconducting Single-Photon Detectors for Dark Matter Search}

\Author{Federico Paolucci $^{1,\dagger}$ and Francesco Giazotto$^{1,\ddagger}$}

\AuthorNames{Firstname Lastname, Firstname Lastname and Firstname Lastname}

\address{$^{1}$ \quad NEST, Istituto Nanoscienze-CNR and Scuola Normale Superiore, I-56127 Pisa, Italy}

\firstnote{federico.paolucci@nano.cnr.it} 
\secondnote{francesco.giazotto@sns.it}

\abstract{The composition of dark matter is one of the puzzling topics in astrophysics. 
To address this issue, several experiments searching for the existence of axions have been designed, built and realized in the last twenty years.
Among all the others, light shining through walls experiments promise to push the exclusion limits to lower energies. For this reason, effort is put for the development of single-photon detectors operating at frequencies $<100$ GHz. Here, we review recent advancements in superconducting single-photon detection. In particular, we present two sensors based on one-dimensional Josephson junctions with the capability to be in situ tuned by simple current bias: the nanoscale transition edge sensor (nano-TES) and the Josephson escape sensor (JES). These two sensors are the ideal candidates for the realization of microwave light shining through walls (LSW) experiments, since they show unprecedented frequency resolutions of about 100 GHz and 2 GHz for the nano-TES and JES, respectively.}


\begin{document}


\section{Introduction}
\begin{figure}[t!]
\centering
\includegraphics[width=0.9\columnwidth]{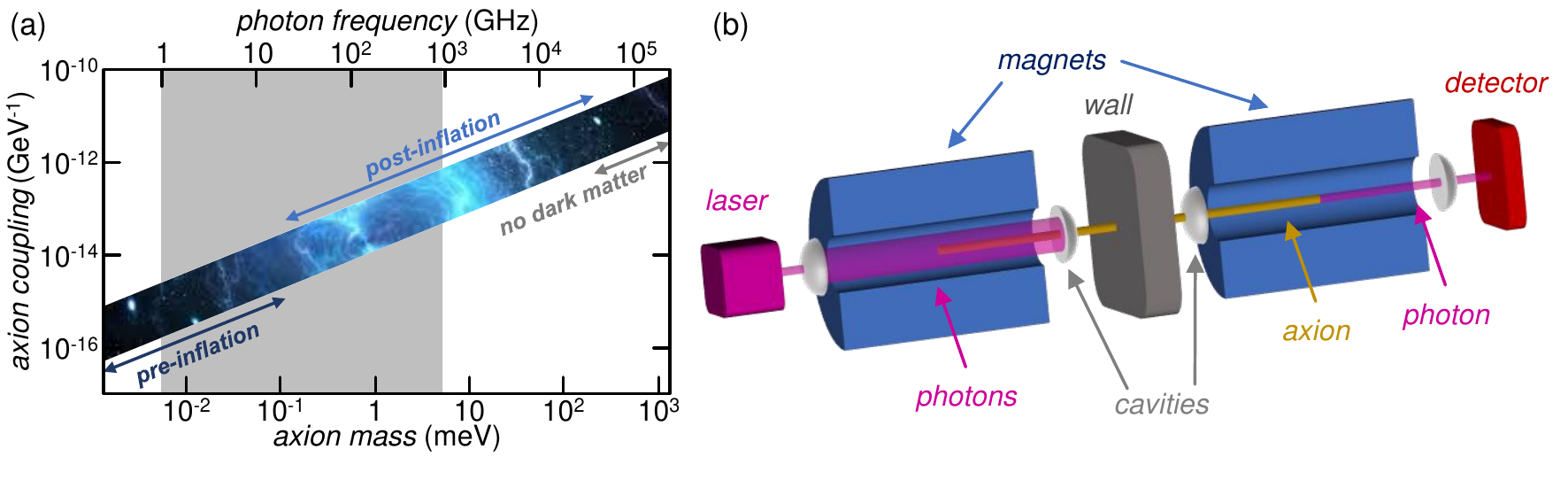}
\caption{Laboratory axion search. (\textbf{a}) Photon-axion coupling ($g$) versus axion mass, where axions origins are indicated. The diagonal band shows the parameter space consistent with the quantum chromodynamics (QCD) axion from the Peccei-Quinn theory. The grey area depicts the operating range of the detectors presented in this review. (\textbf{a}) Conceptual representation of a light-shining-through-wall (LSW) experiment. A laser (violet) feeds photons in a Fabry-Perot cavity immersed in a constant magnetic field. An axion is generated by the conversion of a photon through Primakoff effect and passes through the wall (grey). The axions converts back in a photon of same energy in the second magn etic field area and is revealed by a single-photon detector (red).}
\label{FigAX}
\end{figure}

Axions and weakly interacting massive particles (WIMPs) are expected to be possible candidates of cold dark matter \cite{Preskill,Abbott}. Furthermore, axions and axion-like particles (ALPs) are proposed to solve the charge-conjugation parity (CP) problem in quantum chromodynamics (QCD) by means of the Peccei-Quinn mechanism \cite{Peccei, Dine, Kim}. Up to now, the experimental searches focusing on axions or ALPs have produced null results with corresponding excluded regions in the coupling constant ($g$) versus mass parameter space shown in Fig. \ref{FigAX}(a). Two classes of experiments are performed: astrophysical experiments observing astrophysical phenomena or attempting to detect cosmic axions, and laboratory-based experiments which aim to demonstrate the existence of axions in strictly controlled settings \cite{Zioutas}. 

The design of experiments based on the observation of solar axions is strongly affected by the limits of the solar models. Indeed, the coupling of low-mass weakly interacting particles produced in the sun with normal matter is bounded by the observations of stellar lifetimes and energy loss rates. Solar models together with measurement of neutrino fluxes constraints the magnitude of the coupling constant to $g \leq 7\times 10^{-10}$ GeV$^{-1}$. In addition, the presence of ALPs created by the Primakoff process \cite{Sikivie, Bibber}, consisting in photon-axion conversion in an external magnetic field, would alter stellar-evolution.
Instead of using stellar energy losses to infer the axion exclusion limits, the flux of axions created by the sun can be detected through axion helioscopes, such as CAST \cite{Arik} and IAXO \cite{Armengaud} experiments. These experiments constantly point at the sun by means of a tracking system aiming at converting the solar axions into detectable X-ray photons through Primakoff effect. In microwave cavity experiments, such as ADMX \cite{Asztalos} and QUAX \cite{Barbieri}, galactic halo axions may be detected by their resonant conversion into a quasi-monochromatic microwave signal in a high-quality-factor electromagnetic cavity permeated by a strong static magnetic field. The resonance frequency of the cavity is tuned to equalize the total axion energy. Interestingly, only these experiments are able to probe part of the QCD Peccei-Quinn region \cite{Graham}.

Light shining through walls (LSW) experiments, such as ALPS \cite{Asztalos} and STAX \cite{Capparelli}, are searching techniques based on the laboratory creation and detection of axions. The general concept of a LSW experiment is shown in Fig. \ref{FigAX}(b). A laser beam is sent through a long magnet, allowing for the coherent photon-axion conversion due to the Primakoff effect. The wall acts as photon barrier, thus blocking the laser beam, while allowing axions to pass through due to its almost zero crosssection for interactions with baryonic matter. A second magnet placed after the wall causes the photon-axion back conversion. Since both conversions are very rare (depending on $g^4$), very intense sources are necessary. The highest luminosity photon sources currently available are the gyrotrons, that operate typically below the THz region, with a maximum power of 1 MW at about 100 GHz. In this spectral region, single-photon detection is extremely difficult. In fact, LSW experiments at the microwave have been proposed \cite{Capparelli} but not realized yet. 

To implement microwave LSW experiments, the key ingredient is thus the development of new ultrasensitive single-photon detectors operating at unprecedented low frequencies $f\leq100$ GHz \cite{Alesini1, Alesini2}. Nowadays, state of the art detectors for astrophysics are mainly based on transition edge sensors (TESs) \cite{Irwin1995,Irwin2006} and kinetic inductance detectors (KIDs) \cite{Khosropanah,Visser,Monfardini}. A strong reduction of the thermal exchanges in the sensing elements is fundamental to push single-photon detection to lower frequency (energy), since their operation principle relies on the change of the electronic temperature due to photon absorption. To improve the sensitivity of superconducting detectors, miniaturization and Josephson effect \cite{Josephson} have been exploited \cite{Solinas,Guarcello,Virtanen,Giazotto2008}. In particular, single-photon detectors based on superconducting nanowires, the superconducting nanowire single-photon detectors (SNSPDs), have been developed in the visible and infrared bands \cite{Natarajan}. Furthermore, single-photon counters based on tunnel Josephson junctions \cite{Oelsner, Oelsner2} have been proposed for the detection of axions in the GHz range \cite{Kuzmin}.

Recently, two microwave single-photon detectors, nanoscale transition edge sensors (nano-TES) \cite{Paolucci2} and the Josephson escape sensor (JES) \cite{Paolucci}, have been designed. These devices are based on the use of a one dimensional fully superconducting Josephson junction (1DJ) as radiation absorber. The nano-TES and the JES point towards unprecedented frequency resolutions of about 2 GHz thus enabling the possibility to implement LSW experiments. In addition, differently from all the other superconducting detectors the sensitivity of these sensors can be in situ tuned by simple current biasing. 

This paper reviews these sensors in regards of their theoretical and experimental properties. In particular, Section \ref{TheoryJJ} presents the theoretical description of the 1DJs, while Sec. \ref{Exp1DJ} shows their experimental electronic and thermal transport properties. Section \ref{Theory} introduces the operation principles of the nano-TEs and JES detectors. Section \ref{ExpPerf} presents the detection performance of the nano-TES and the JES.
Section \ref{Mater} presents the experimental methods used for the sensors characterization. Finally, Sec. \ref{Concl} summarizes the results and opens for new applications for nano-TES and JES detectors.

\section{Theoretical modelling of a one-dimensional fully superconducting Josephson junction}
\label{TheoryJJ}

A Josephson junction (JJ) is a structure where the capability of a superconductor to carry a dissipationless current is strongly suppressed. Typically, the discontinuity of the supercurrent flow is realized by separating the two superconducting elements by means of a \emph{weak link}. A weak link can be realized through a variety of structures. The most common realizations are 
\begin{itemize}
    \item a thin insulating barrier forming a superconductor/insulator/superconductor SIS-JJ;
    \item a short section of normal metal creating superconductor/normal metal/superconductor SNS-JJ;
    \item a physical constriction in the superconductor producing an SsS-JJ (known as Dayem bridge);
    \item a short section of lower energy gap superconductor realizing a SS'S-JJ.
\end{itemize}

\begin{figure}[t!]
\centering
\includegraphics[width=0.9\columnwidth]{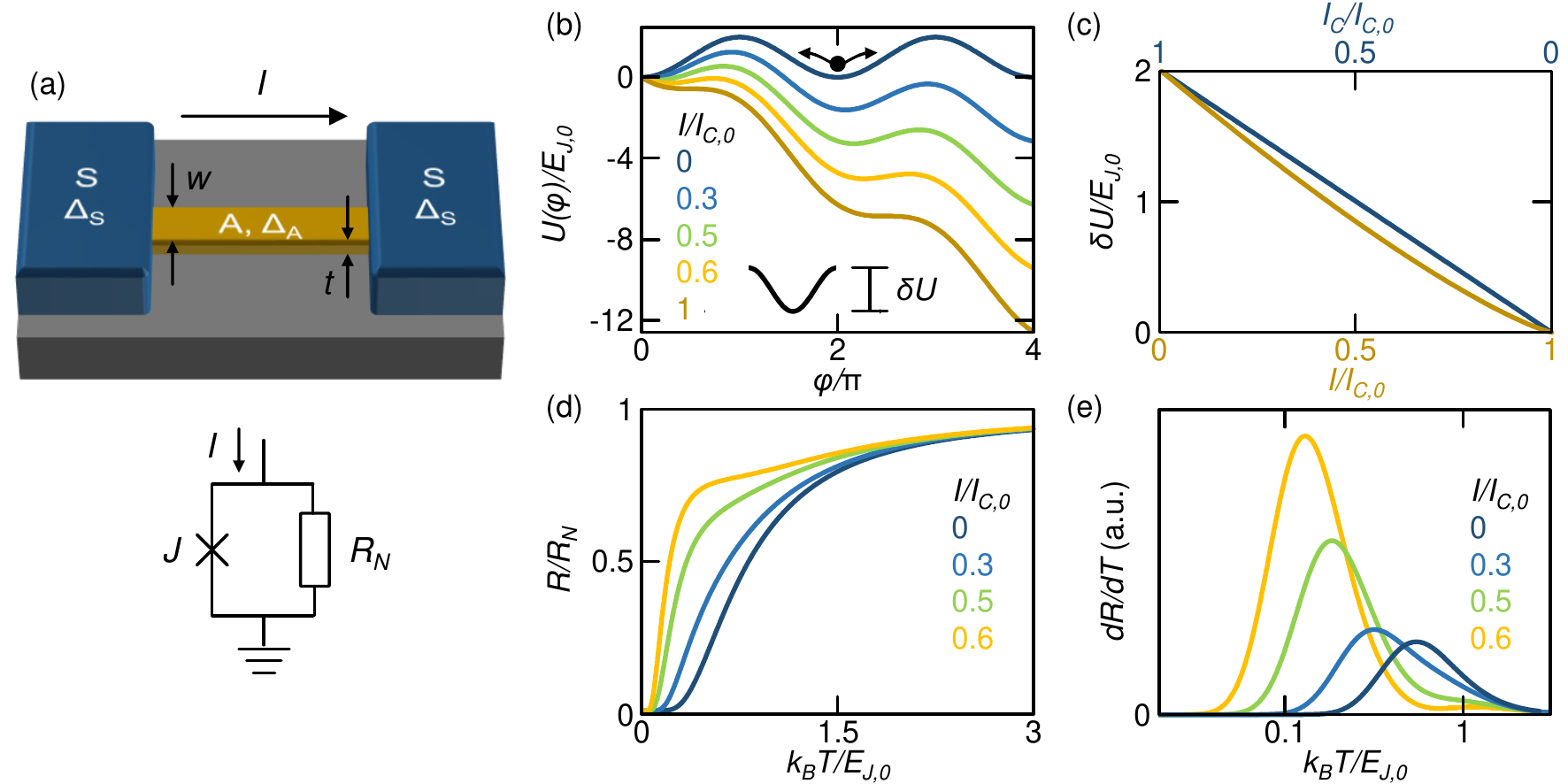}
\caption{\textbf{Structure and transport properties of a 1DJ.} (\textbf{a}) Top: Scheme of the structure a 1DJ. Two superconducting electrodes ($S$, blue) are separated by a weak link consisting of a superconducting wire ($A$, bronze). Its width ($w$) and thickness ($t$) are indicated. The two superconducting energy gaps follow $\Delta_A\ll \Delta_S$. The current ($I$) flowing along the 1DJ is shown. Bottom: RSJ model of a 1DJ where $I$ is the bias current, $J$ is the junction and $R_N$ is the shunt resistor. (\textbf{b}) Tilted washboard potential of a 1DJ calculated for different values of $I$.  The energy  barrier ($\delta U$) for the escape of the phase particle from the washboard potential (WP) decreases by rising the bias current, thus the probability of the transition of the 1DJ to the normal-state increases with $I$. The phase particle in the WP \cite{tinkham} is indicated by means of the black dot. (\textbf{c}) Energy barrier ($\delta U$) normalized with respect to the zero-temperature Josephson energy ($E_{J,0}$) calculated by varying the critical current (top) and the bias current (bottom), respectively. (\textbf{d}) temperature dependence of the normalized resistance ($R/R_N$) calculated for selected values of $I$. $R_N$ is the normal-state resistance of the 1DJ. (\textbf{e}) Temperature derivative of the resistance ($\text{d}R/\text{d}T$) calculated for the same values of $I$ in panel (d).}
\label{FigJJ}
\end{figure}

Here, we focus at one-dimensional SS'S-JJs, that we will call 1DJ, where the two superconducting lateral electrodes ($S$) are separated by a one-dimensional wire ($A$) made of a different superconductor. In a one-dimensional superconductor, both the thickness ($t$) and width ($w$) of $A$ are smaller than the London penetration depth ($\lambda_{L,A}$) and the Cooper pairs coherence length ($\xi_A$).
This ensures uniform superconducting properties of $A$ along its cross section. In particular, a one-dimensional superconductor shows a constant superconducting wave function, a homogeneous supercurrent density, and uniform penetration of $A$ by an out-of-plane magnetic field.
In the following, we will name such a structure as 1DJ for simplicity. The general structure of a 1DJ is shown in Fig. \ref{FigJJ}(a).

The behavior of the 1DJ can be described through the overdamped resistively shunted junction (RSJ) model \cite{tinkham}, where the JJ is shunted by its normal-state resistance [see Fig. \ref{FigJJ}(a)]. The bias current ($I$) dependence on the stochastic phase difference [$\varphi(t)$] over the junction is
\begin{equation}
\frac{2e}{\hbar} \frac{\dot{\varphi}(t)}{R_N} + I_C \sin{\varphi(t)} = I + \delta I_{th}(t) \text{,}
\label{CPR}
\end{equation}
where $e$ is the electron charge, $\hbar$ is the reduced Planck constant, $R_N$ is the wire normal-state resistance, while $I_C$ is its critical current. The normal-state resistance of the 1DJ acts as shunt resistor providing a thermal noise contribution to the flowing current given by $\left\langle \delta I_{th}(t)\delta I_{th}(t')\right\rangle=\frac{k_BT}{R_N}\delta{(t-t')}$, where $k_B$ is the Boltzmann constant and $T$ is the temperature. The transition to the normal-state of a JJ or a superconducting nano-wire is usually attributed $2\pi$ quasiparticle phase-slips \cite{Barone1982,tinkham}, because a full phase rotation entails passing through the condition $I_C=0$. Within the RSJ model, the phase-slip is the motion of a phase particle in a tilted washboard potential (WP) under the presence of friction forces. 
The WP can be written
\begin{equation}
U(\varphi)=-\frac{\hbar I}{2e}\varphi-\delta U \cos{\varphi}
\text{,}
\label{WP}
\end{equation}
where $\delta U=\delta U(I,E_J)$ is the escape energy for the phase particle. It is worth to note that the only parameter dependent on the JJ geometry is $\delta U(I,E_J)$. According to \cite{Bezryadin2012}, the $\delta U(I, E_j)$ can be parametrized as follows:
\begin{equation}
\delta U(I,E_J)\sim 2 E_J\left(1-I/I_C \right)^{5/4}=\frac{\Phi_0 I_C}{\pi}\left(1-I/I_C \right)^{5/4}\text{.}
\label{eq:potential}
\end{equation}
Equations \ref{WP} and \ref{eq:potential} show that both bias current and Josephson energy ($E_J=\Phi_0I_C/2\pi$ with $\Phi_0\simeq 2.067\times 10^{-15}$ Wb the flux quantum) define the WP. In particular, $\delta U$ is suppressed by lowering the Josephson energy and rising the bias current. The latter also produces the tilting of the WP, as shown in Fig. \ref{FigJJ}(b). It is interesting to quantitatively compare the effects of $I$ and $E_J$ on the WP. To provide this comparison, in Eq. \ref{eq:potential} we replaced the Josephson energy with its critical current dependent relation. With no current bias ($I=0$), the barrier depends linearly on the critical current. Instead, the bias current has a stronger impact on $\delta U$, since $\delta U\sim I_C^{-5/4}$. The comparison between the two methods to suppress the energy barrier is shown in Fig. \ref{FigJJ}(c). Thus, the current bias is the most efficient method to control the supercurrent flowing in a 1DJ. 

The normal-state resistance of a 1DJ is low in comparison with the sub-gap resistance (and the case of a tunnel Josephson junction). Since both the normal-state resistance and the capacitance of the junction are small, the Stewart-McCumber parameter obeys to $\beta_C\ll1$. Therefore, the 1DJ can be described by means of the overdamped junction limit of the RSJ model. \cite{tinkham} In this approximation, the temperature dependence of the voltage drop build across a 1DJ can be written \cite{Ivanchenko1969} 
\begin{equation}
V(I, E_J, T)=R_N \left[ I - I_{C,0} \operatorname{Im}\frac{\mathcal{I}_{1-iz}\left( \frac{E_J}{k_BT}\right) }{\mathcal{I}_{-iz}\left( \frac{E_J}{k_BT}\right)}\right]\text{,}
\label{eq:voltage}
\end{equation}
where $I_{C,0}$ is the junction zero-temperature critical current, $\mathcal{I}_{\mu}(x)$ is the modified Bessel function with imaginary argument $\mu$, and the imaginary argument takes the form $z=\frac{E_J}{k_BT}\frac{I}{I_C}$. Therefore, $V$ strongly depends on $I_C$ (thus $E_J$) and $I$. The current derivative of the voltage drop calculated at different values of temperature provides the $R(T)$ characteristics
\begin{equation}
R(I, E_J, T)=\frac{\text{d}V(I, E_J, T)}{\text{d}I}\text{.}
\label{resistance}
\end{equation}

By solving Eq. \ref{resistance} for different values of $I$, we can evaluate the impact of the bias current on the resistance versus temperature characteristics of ta 1DJ. In particular, Fig. \ref{FigJJ}(d) highlights that the temperature of the superconducting-to-resistive state transition of the JJ decreases by  rising $I$. Furthermore, high values of bias current have a second important effect on the $R(T)$: the temperature span suitable for the superconducting-to-normal-state transition becomes narrower. The temperature derivative of $R(T)$ confirms the positive impact the current bias on the transition width, as shown by Fig. \ref{FigJJ}(e). This behavior is related to the decrease of $\delta U$ and to the current induced tilting of the WP (providing a preferred direction of the phase-slips). 

\begin{figure}[t!]
\centering
\includegraphics[width=0.9\columnwidth]{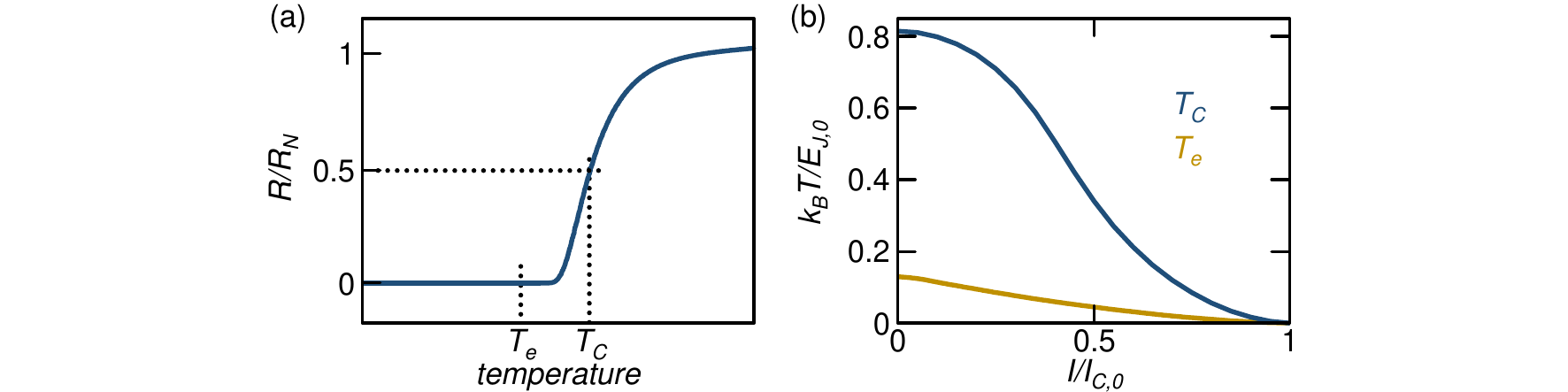}
\caption{\textbf{Definition of escape and critical temperature.} (\textbf{a}) Calculated normalized resistance ($R/R_N$) versus temperature calculated for a 1DJ within the RSJ model. The escape temperature ($T_e$) and the critical temperature ($T_C$) are indicated. (\textbf{b}) Critical temperature ($T_C$, blue) and escape temperature ($T_e$, orange) as a function of the bias current ($I$) calculated by means of the RSJ model of a 1DJ.}
\label{FigTdef}
\end{figure} 

Two characteristic temperatures related to the superconductor-to-normal-state transition can be defined, as shown in Fig. \ref{FigTdef}(a). On the one hand, the effective critical temperature ($T_C$) is the temperature corresponding to half of the normal-state resistance [$R(T_C)=R_N/2$]. On the other hand, the escape temperature ($T_e$) is the highest temperature providing a zero resistance of the 1DJ [$R(T_e)=0$]. The bias current has a strong influence on both $T_C$ and $T_e$, as shown in Fig. \ref{FigTdef}(b). In particular, the effective critical temperature decreases much faster than the escape temperature by increasing $I$, $T_C\sim T_e$ for $I\to I_C$. As a consequence, the superconducting-to-normal-state transition becomes sharper.

\section{Experimental demonstration of a 1DJ}
\label{Exp1DJ}
This section is dedicated to the experimental demonstration of bias current tuning of the $R$ versus $T$ characteristics of a 1DJ. In particular, Sec. \ref{DensExp} aims to proof that the structure under study is one-dimensional, while Sec. \ref{TranExp} will show the bias current control of the superconducting-to-normal phase transition.

\subsection{Density of states and one-dimensionality}
\label{DensExp}
\begin{figure}[t!]
\centering
\includegraphics[width=0.9\columnwidth]{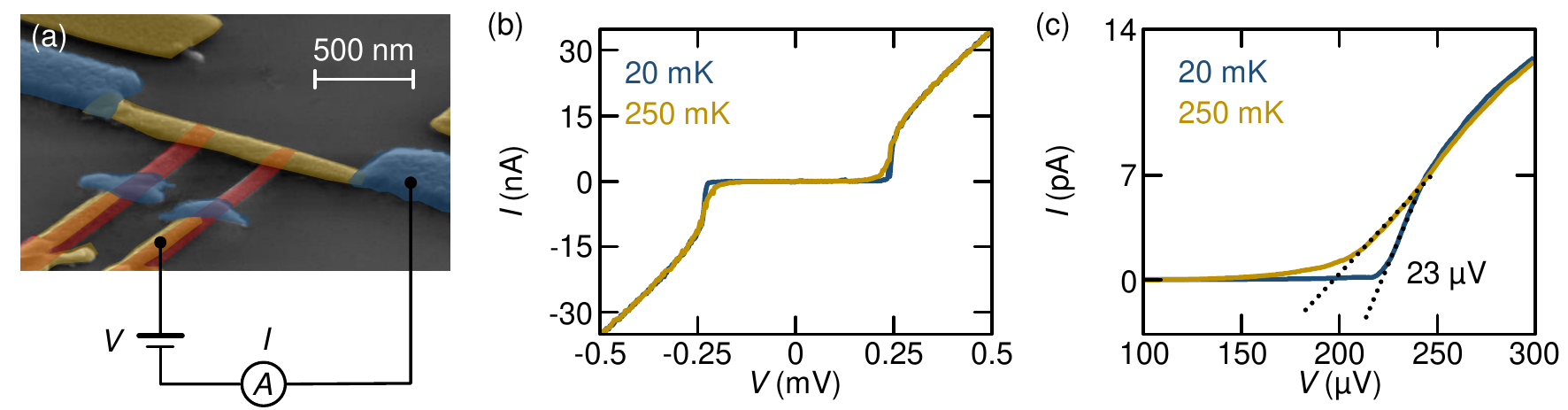}
\caption{\textbf{Measurement of the density of states in a 1DJ.} (\textbf{a}) False-color scanning electron micrograph of a device used to measure the DOS of a 1DJ. The 1DJ is made of an Al/Cu bilayer nanowire (yellow) interrupting two Al electrodes (blue). The Al probes (red) allow to perform tunnel spectroscopy. In the experiment, a voltage ($V$) is applied between $A$ and one probe while recording the current ($I$). (\textbf{b}) Tunneling current ($I$) as a function of voltage ($V$) characteristics recorded at $T_{bath}=20$ mK (blue) and $T_{bath}=250$ mK (yellow). (\textbf{c}) Zoom of the $IV$ characteristics in correspondence of the transition to the normal-state. It is possible to extract $\Delta_{A,0}\simeq23\;\mu$eV and $\Delta_{P,0}\simeq200\;\mu$eV as the crossing between the black dotted lines and $I=0$.}
\label{FigGT}
\end{figure}

A typical 1DJ is realized in the form of a 1.5 $\mu$m-long ($l$), 100 nm-wide ($w$) and 25 nm-thick ($t$) Al/Cu bilayer nanowire-like active region sandwiched by two Al electrodes. The detailed fabrication procedure is described in Sec. \ref{Mater}. To ensure that the JJ is one-dimensional ($\xi_A > t,w$ and $\lambda_{L,A} > t,w$), a full spectroscopic characterization of the active region ($A$) is necessary. To measure the density of states of the active region, the test device is equipped with two additional Al tunnel probes, as shown by the false-color scanning electron micrograph (SEM) in Fig. \ref{FigGT}(a). The $IV$ tunnel characteristics of $A$ are obtained by applying a voltage ($V$) and measuring the current ($I$) flowing between one lateral electrode and a tunnel probe (that we indicate with $P$). The experimental set-up is described in detail in Sec. \ref{Mater}.

The energy gap of a superconductor is temperature independent up to $T\sim 0.4\;T_C$ thus implying $\Delta(T)=\Delta_0$, with $\Delta_0$ its zero-temperature value \cite{tinkham}. Since aluminum thin films typically show a $T_C\geq 1.2$ K \cite{cochran}, the superconducting gap of the aluminum probes is temperature independent up to at least 500 mK. In this temperature range the energy gap of the nanowire is strongly temperature dependent, since the inverse proximity effect weakens its superconducting properties (it is a superconductor/normal metal bilayer). As a consequence, this experimental set-up can be used to study the superconducting properties of $A$. In particular, the zero-temperature energy gap ($\Delta_{0,A}$) will be helpful to demonstrate the one-dimensionality of the nanowire.

To obtain $\Delta_{0,A}$, the $IV$ characteristics were measured at base temperature ($T=20$ mK) and well above the expected critical temperature of $A$, but below $0.4\;T_{C,Al}$ ($T=250$ mK), as shown in Fig. \ref{FigGT}(b). At the base temperature, both $A$ and $P$ are in the superconducting state. Therefore, the voltage bias needs to reach $V=\pm(\Delta_{A,0}+\Delta_{P,0})/e$ (with $\Delta_{0,P}$ the zero-temperature gap the Al probe) to switch to the normal-state \cite{Giazotto2006}. On the contrary, at $T_{bath}= 250$ mK the nanowire is in the normal-state thus the transition occurs at $V=\pm\Delta_{P,0}/e$. The resulting zero-temperature energy gap of the Al probe is  $\Delta_{0,P}\simeq200\;\mu$eV [see Fig. \ref{FigGT}(c)], therefore indicating a critical temperature $T_{C,P}=\Delta_{P,0}/(1.764k_B)\simeq 1.3$ K. Furthermore, the difference between the onset of the rise of the tunnel current between the curves recorded at 20 mK and and 250 mK provides $\Delta_{A,0}\simeq23\;\mu$eV thus indicating a critical temperature $T_{C,A}\simeq150$ mK.

A 1DJ requires the intrinsic superconducting properties of the nanowire to be uniform and dominate over the proximity effect induced by the lateral banks. 
The latter could induce an energy gap in a non-superconducting Al/Cu bilayer given by $E_g\simeq3\hbar D_A/l^2\simeq5\;\mu$eV \cite{Golubov}, where $D_A$ is the diffusion constant of the active region. The details are given in the Methods section.
Since $E_g\sim 0.25\Delta_{A,0}$, the superconducting properties of $A$ are dominated by the intrinsic superconductivity of the Al/Cu bilayer.

If the Al/Cu bilayer lies in the Cooper limit \cite{DeGennes1964,Kogan1982}, it can be considered a uniform superconductor. The Cooper limit has two requirements: negligible contact resistance between the two layers and thickness of each layer lower than its coherence length. Since its large surface area, the Al/Cu interface resistance is negligibly small in comparison with the nanowire normal-state resistance, thus fulfilling the first requirement. In addition, the superconducting Al film fulfils $\xi_{Al}\simeq80$ nm $\gg t_{Al}=10.5\;\text{nm}$. At the same time, the Cu layer obeys to $\xi_{Cu}=\simeq255$ nm $\gg t_{Cu}=15\;\text{nm}$. Therefore, the second condition is fulfilled, too. We can conclude that the Al/Cu bilayer respects the Cooper limit and $A$ can be considered as formed from a single superconducting material. The details are given in the Methods section.

We can now discuss the one-dimensionality of $A$. In particular, the superconducting coherence length in $A$ is $\xi_A\simeq 220$ nm, Since this value is much larger than its thickness ($\xi_A\simeq 220$ nm$\gg t=t_{Al}+t_{Cu}=25.5\;\text{nm}$), the pairing potential of the bilayer is constant along the z axis. Furthermore, the active region is one-dimensional with respect to the superconducting coherence length, because $\xi_A\gg w=100$ nm. Since the London penetration depth of $A$ is $\lambda_{L,A}\simeq 970$ nm, the nanowire is 1D with respect to the London penetration depth, since $\lambda_{L,A}\gg t,w$.

In conclusion, we have demonstrated that the Al/Cu bilayer embedded between two Al electrodes forms a 1DJ. Therefore, this structure can be used to investigate the impact of $I$ on the $R(T)$ characteristics.

\begin{figure}[t!]
\centering
\includegraphics[width=0.9\columnwidth]{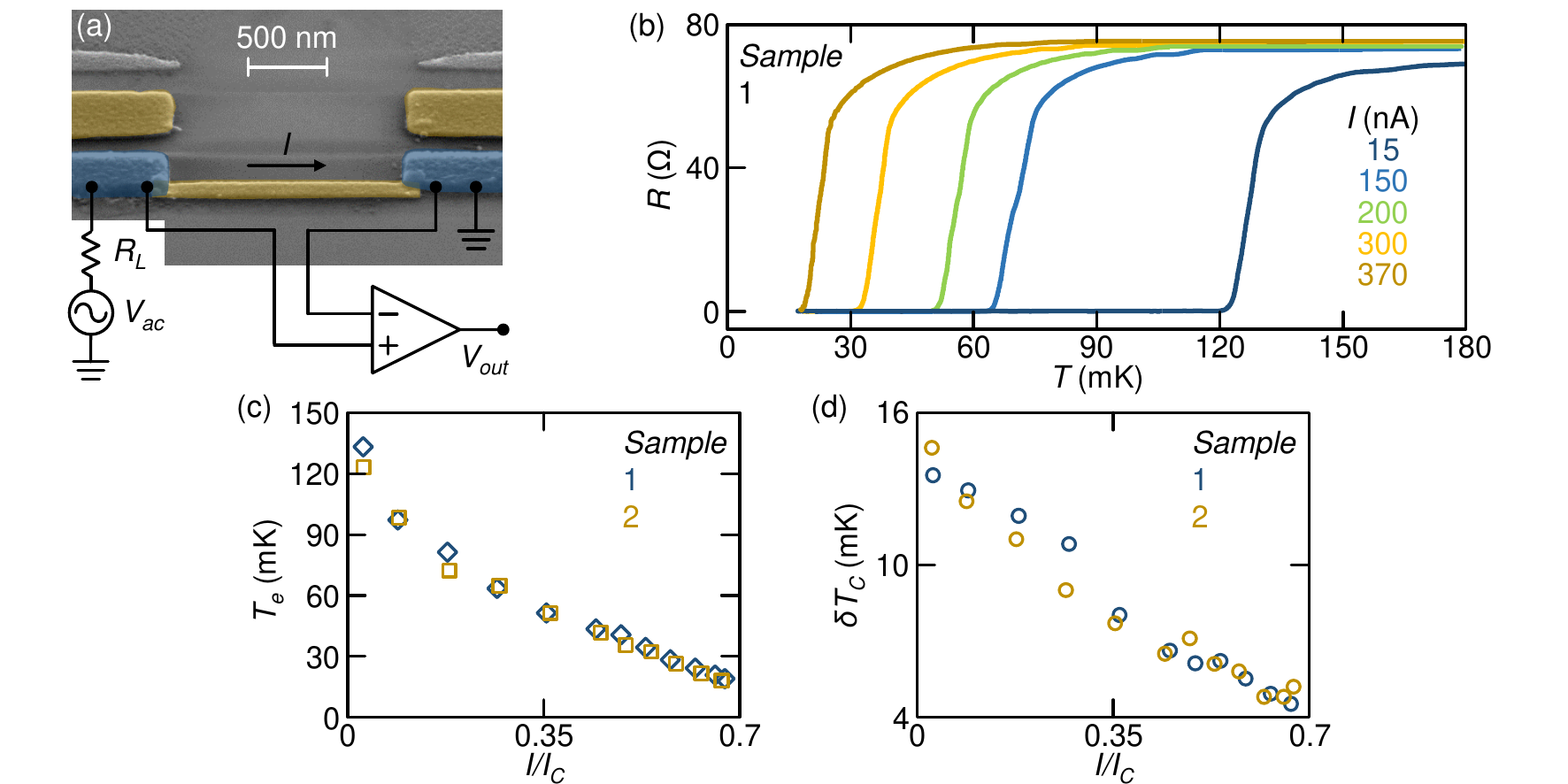}
\caption{\textbf{Current control of the resistance versus temperature characteristics in a 1DJ.} (\textbf{a}) False-colored scanning electron micrograph of a 1DJ. The nanowire is made of an Al/Cu bilayer (yellow) separating two thick Al electrodes (blue).
The 1DJ is AC current-biased (amplitude $I$), while the voltage drop across the wire ($V_{out}$) is measured through a lock-in amplifier. The load resistor ($R_L \gg R_N$) guarantees constant bias current while transitioning to the normal-state. (\textbf{b}) Selected normalized resistance ($R$) versus temperature ($T$) characteristics recorded for different values of bias current ($I$). (\textbf{c}) Dependence of the escape temperature ($T_e$) on bias current normalized with respect to the zero-temperature 1DJ critical current ($I/I_{C}$) for two different samples. (\textbf{b}) Width of the phase transition ($\delta T_C$) versus $I$ for two different 1DJs.}
\label{FigImod}
\end{figure}

\subsection{Current control of the R vs T}
\label{TranExp}

To investigate the impact of the bias current on the transport properties of a 1DJ, the resistance $R$ vs temperature characteristics were obtained by conventional four-wire low-frequency lock-in technique by varying the excitation current amplitude from 15 nA to 370 nA . The current was generated by applying a voltage ($V_{ac}$) to a load resistor ($R_L$) of impedance larger than the device resistance ($R_L=100$ k$\Omega\gg R_N\simeq 77\; \Omega$), as shown in Fig. \ref{FigImod}(a). For the details regarding the device fabrication and experimental set-up see Sec. \ref{Mater}.

The magnetic field generated at the wire surface in correspondence of the maximum bias current is $B_{I,max}\simeq 4.7$ $\mu$T. This value is orders of magnitude lower than the critical magnetic field of $A$ that was measured to be about 21 mT \cite{Paolucci}. So, the self-generated magnetic field does not affect the properties of the 1DJ.

The resistance versus temperature characteristics shift towards low temperatures by rising the current from $\sim 3\%$ and $\sim 65\%$ of $I_{C,0}$. In addition, the $R(T)$ characteristics preserve the same shape up to the highest bias currents. The use of an AC bias allowed to resolve the $R$ vs $T$ characteristics near the critical temperature. In fact, values of DC bias higher than the retrapping current \cite{Courtois} ($I_R$, that is the switching current from the resistive to the dissipationless state) would cause the sudden transition of the device resistance to $R_N$. Instead, the AC bias has always a part of the period lower than $I_R$ thus enabling the precise measurement of the entire $R(T)$ traces.

The electronic temperature of the nanowire ($T_A$) at the middle of the phase transition under current injection is different from $T_{bath}$, since Joule dissipation (for $R\ne0$) causes the quasiparticles overheating in $A$ yielding $T_A > T_{bath}$ \cite{Giazotto2006}. Therefore, from the $R$ vs $T$ curves we can only investigate the current-dependent escape temperature [$T_e(I)$]. The values of $T_e$ are shown in Fig. \ref{FigImod}(c) as a function of $I/I_C$ for two different samples. The escape temperature is monotonically reduced by rising the bias current with a minimum value $\sim20$ mK for $I=370$ nA, that is $\sim 15\%$ of the intrinsic critical temperature of the active region, $T_C^i\sim 130$ mK.

The width of the superconducting-to-normal-state transition ($\delta T_C$) reduces by increasing the current injection, as shown in Fig. \ref{FigImod}(d). In particular, $\delta T_C$ is suppressed by a factor of $4$ at the largest value of bias current. It is worth mentioning that this behavior is in full agreement with the theoretical behavior of a 1DJ shown in Sec. \ref{TheoryJJ}. Therefore, in the following we will focus on the detection properties of a 1DJ.

\begin{figure}[t!]
\centering
\includegraphics[width=\columnwidth]{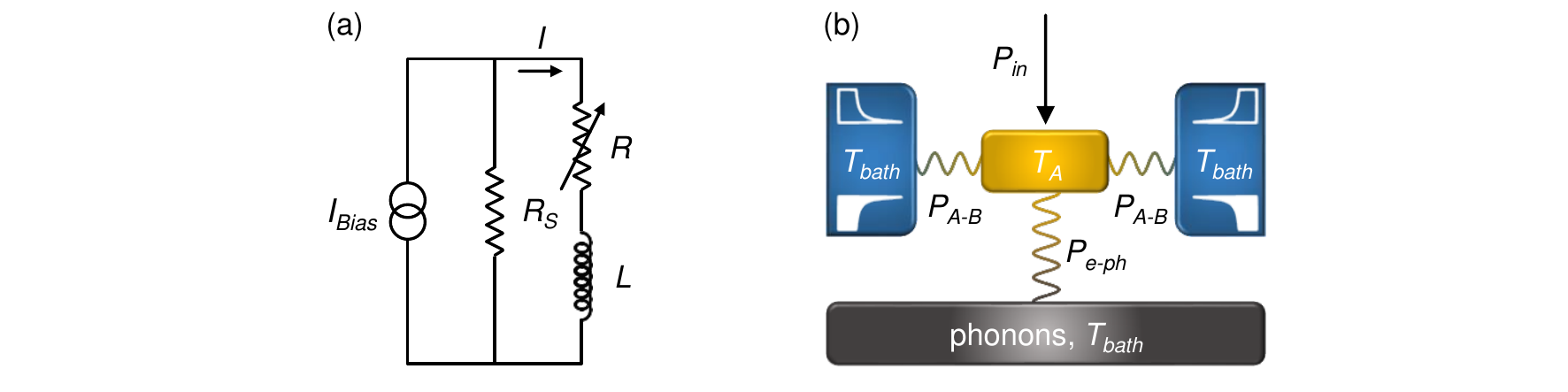}
\caption{\textbf{General thermal and electrical model of the nano-TES and JES.} (\textbf{a}) Schematic representation of a typical biasing circuit for the nano-TES and JES. The parallel connection of the sensor (of variable resistance $R$) and the shunt resistor ($R_{S}$) is biased by the current $I_{Bias}$. The role of $R_S$ is to limit the Joule overheating of $A$ when transitioning to the normal-state. The variations of the current ($I$) flowing through the sensing element are measured by means of the inductance $L$. For instance, an inductively-coupled SQUID amplifier could serve as read-out element. (\textbf{b}) Thermal model of the nano-sensors where the main thermal exchange channels are shown. 
$P_{in}$ is the power coming from the incoming radiation, $P_{e-ph}$ is the heat exchanged between electrons in the active region (yellow) at $T_A$ and lattice phonons (grey) at $T_{bath}$, while $P_{A-B}$ is the heat current flowing towards the superconducting electrodes (blue) residing at $T_{bath}$.}
\label{FigHE}
\end{figure}

\section {Operation principle of the nano-TES and JES}
 \label{Theory}
The 1DJ was used to design two single-photon detectors operating in the GHz band: the nanoscale transition edge sensor (nano-TES) \cite{Paolucci2} and the Josephson escape sensor (JES) \cite{Paolucci}. These sensors take advantage of the strong resistance variation of the superconducting nanowire while transitioning to the normal-state, such as in a conventional TES \cite{Irwin1995} and SNSPDs \cite{Natarajan}. Differently from all the other superconducting radiation detectors, the sensitivity of the nano-TES and the JES can be in situ controlled, since the resistance versus temperature characteristics of a 1DJ can be tuned by varying the bias current. As a consequence, the 1DJ serves as the active region of these sensors. The main difference between the nano-TES and the JES is the operating temperature. Indeed, the nano-TES operates at $T_C$, i.e., at the middle of the superconductor-to-normal-state transition [see Fig. \ref{FigTdef}(a)], while the JES operates at $T_e$, i.e., deeply in the superconducting state. Notably, these temperatures can be very different at large bias currents [see Fig. \ref{FigTdef}(b)].

For both sensors, the absorption of radiation triggers an increase of the electronic temperature in the superconducting nanowire ($T_A$) thus driving its transition to the normal-state. The latter would generate Joule heating in the active region when biased with a constant current with consequent thermal instability. To solve this issue, the nano-TES and the JES could be biased with the circuitry shown in Fig \ref{FigHE}(a). The shunt resistor ($R_S$) limits the current ($I$) flowing through the sensor ($R$) when the $A$ undergoes the superconducting-to-normal-state transition. This is called  negative electrothermal feedback (NEFT) \cite{Irwin1995}. For the nano-TES, the sensor is biased at $T_C$ ($R=R_N/2$), therefore the condition for the shunting resistor reads $R_S=IR_N/[2(I_{Bias}-I)]$, where $I_{Bias}$ is the current provided by the generator. For the JES, the device is operated at $T_e(I)$,i.e. at $R=0$, and the role of $R_S$ is to limit the current flow through the sensing element below $I_R$. This happens for $R_S\leq R_NI_R/I_{bias}$ and brings $A$ quickly back to the superconducting state after radiation absorption. Therefore, the sensor always operates in the superconducting state. For both the nano-TES and the JES, the variations of $I$, due radiation absorption, can be measured via a conventional SQUID amplifier coupled to the inductance $L$ \cite{Paolucci2}.

The ability of a superconducting sensor to resolve a single-photon depends on its ability to convert the power of the incoming radiation into a change of electronic temperature in the active region. The latter is related to the predominant thermal exchange mechanisms occurring in $A$. Figure \ref{FigHE}(b) shows the thermal model describing the active region of both the nano-TES and the JES. Here, $P_{in}$ is the power released into the active region by the external radiation, $P_{e\text{-}ph}$ is the heat thermalization of the quasiparticles with the lattice phonons, and $P_{A\text{-}B}$ represents the energy out-diffusion from the active region to the lateral superconducting leads. When the critical temperature of the lateral electrodes ($T_{C,B}$) is much higher than operating temperature ($T_C$ for the nano-TES and $T_e$ for the JES), they behave as energy filters, the so-called Andreev mirrors \cite{Andreev1964}, thus ensuring perfect thermal insulation of $A$ ($P_{A\text{-}B}\to 0$). Within this condition, $P_{e\text{-}ph}$ is the predominant thermal relaxation channel in the active region. See Ref. \cite{Paolucci2} for the application limits of this assumption.

For the nano-TES, the active region operated almost in the normal-state (at $R_N/2$). Therefore, the electron-phonon coupling of a normal-metal diffusive thin film ca be used \cite{Irwin1995,Giazotto2006}
\begin{equation}
P_{e\text{-}ph,n}=\Sigma_A \mathcal{V}_{A}\left(T_A^5-T_{bath}^5 \right)\text{,}
\label{e-phpowerN}
\end{equation}
where $\mathcal{V}_A$ is the volume of $A$, while $\Sigma_A$ is its electron-phonon coupling constant. The resulting thermal conductance for the active region of a nano-TES ($G_{th,nano\text{-}TES}$) can be calculated through the temperature derivative of the electron-phonon energy relaxation \cite{Irwin1995}
\begin{equation}
G_{th,nano\text{-}TES} = \dfrac{\text{d} P_{e\text{-}ph, n} }{\text{d} T_A} = 5 \Sigma_A {\mathcal{V}}_{A} T_A^4\text{.}
\label{ThermalConductance}
\end{equation}

Differently, the JES operates deeply in the superconducting state at $T_e$ with $A$. Therefore, at very low temperatures the electron-phonon heat exchange is exponentially suppressed with respect to the normal-state \cite{Timofeev2009}
\begin{equation}
P_{e\text{-}ph,s} \propto P_{e\text{-}ph,n} \exp{[-\Delta_A/(k_B T_A)]}\text{,}
\label{e-phpowerS}
\end{equation}
where $\Delta_A$ is the superconducting energy gap in $A$. The thermal conductance of the active region of a JES (operating in the superconducting state) takes the form \cite{Heikkila}

\begin{equation}
G_{th,JES} \approx \dfrac{\Sigma_A \mathcal{V}_A {T_e}^4}{96 \varsigma(5)}
\left[ f_1 \left(\dfrac{1}{\tilde{\Delta}}\right) \cosh(\tilde{h}) e^{-\tilde{\Delta}} + \pi \tilde{\Delta}^5 f_2 (\dfrac{1}{\tilde\Delta})e^{-2 \tilde{\Delta}}  \right] \text{.}
\label{eqGJES}
\end{equation}
The term $f_1$ refers to the electron-phonon scattering, while $f_2$ stems from the recombination processes. In Eq. \ref{eqGJES}, $\varsigma(5)$ is the Riemann zeta function, $\tilde{\Delta} = \Delta_A/k_B T$ is the normalized energy gap of $A$, $\tilde{h} = h/k_B T$ represents exchange field (0 in this case), $f_1(x) = \begin{matrix} \sum_{n=0}^3  C_n x^n \end{matrix} $ with $C_0 \approx 440, C_1 \approx 500, C_2 \approx 1400, C_3 \approx 4700$, and $f_2(x) = \begin{matrix} \sum_{n=0}^2  B_n x^n \end{matrix}$ with $B_0 = 64, B_1 = 144, B_2= 258$. We note that the thermal conductance for a JES is exponentially damped compared to the nano-TES, due to the operation in the superconducting state. Thus, we expect the JES to be extremely more sensitive than a nano-TES operating at the same temperature.

\section{Single-photon detection performance of the nano-TES and the JES.}
\label{ExpPerf}

Microwave LSW experiments for axions search require single-photon detectors of frequency resolution on the order of a few GHz. In the next sections, we will show all the theoretical relations describing the sensing properties of a nano-TES and a JES single-photon detector and the performance inferred from the experimental data.

\subsection{Modelling of the nano-TES}

In order to determine the performances of a sensor in single-photon detection, the frequency resolution is the most used figure of merit, since it defines the lowest energy that the detector can reveal. Indeed, the frequency resolution is related to the thermalization of the quasiparticles in the active region. It identifies the minimum energy able to increase the quasiparticles temperature thus providing a sizeable output signal. For a nano-TES, it can be written \cite{Irwin1995}\\
\begin{equation}
{{\delta\nu}_{TES}} = \dfrac{2.36}{\hbar} \sqrt{\dfrac{4}{\alpha} \sqrt{\dfrac{n}{2}} k_B {T_C}^2 {C_{e,nano-TES}}} \text{,}
\label{FrequencyResolution}
\end{equation}
where $\hbar$ is the reduced Planck constant, $\alpha = \dfrac{\text{d} R}{\text{d} T}\dfrac{T}{R}$ is the electrothermal parameter accounting for sharpness of the phase transition from the superconducting to the normal-state \cite{Irwin1995}, $n=5$ is the electron-phonon coupling exponent for a pure metal and${C_e}_{nano-TES}$ is the electron heat capacitance. It is interesting to note the strongly dependence on $\alpha$ value which determines the NETF mechanism \cite{Irwin1995}. 

Since the nano-TES operates at the critical temperature, the electron heat capacitance of the active region is \cite{Giazotto2006}\\
\begin{equation}
C_{e,nano\text{-}TES} = \gamma_A \mathcal{V}_{A} T_C\text{,}
\label{HeatCapacitance}
\end{equation}
where $\gamma_A$ is the Sommerfeld coefficient of $A$. 

We now focus on the response speed of the nano-TES. By considering the circuit implementing the NETF [see Fig. \ref{FigHE}(b)], the pulse recovery time takes the form \cite{Irwin1995}\\
\begin{equation}
\tau_{eff} = \dfrac{\tau_{nano\text{-}TES}}{1 + \dfrac{\alpha}{n}}\text{,}
\label{taueff}
\end{equation}
where $\tau_{nano\text{-}TES}$ is the intrinsic recovery time of $A$. It can be calculated by solving the time dependent energy balance equation that takes into account all the exchange mechanisms after radiation absorption \cite{Giazotto2006}. The re-thermalization of the quasiparticles to the equilibrium is an exponential function of the time with time constant ($\tau_{nano\text{-}TES}$) given by the ratio between the thermal capacitance and the thermal conductance of $A$
\begin{equation}
\tau_{nano\text{-}TES} = \dfrac{C_{e,nano\text{-}TES}}{G_{th,nano\text{-}TES}}\text{.}
\label{IntrinsicTime}
\end{equation}
Since $\alpha \gg n$, the pulse recovery time is much shorter than the intrinsic time constant of $A$ ($\tau_{eff}<< \tau_{nano\text{-}TES}$). Therefore, the overheating into the active region is decreased by the NETF, thus compensating for the initial temperature variation and avoiding the dissipation through the substrate.

\subsection{Modelling of the JES}

Since the current injection does not change the energy gap of the active region ($\Delta_A\sim \text{const}$), only the effective critical temperature of $A$ changes with $I$, while the intrinsic values of critical temperature ($T_C^i$) is unaffected. As a consequence, being at $T_e(I)$, the JES operates deeply in the superconducting state, thus ensuring high sensitivity (the thermalization is exponentially suppressed by the energy gap, see Eqs. \ref{e-phpowerS} and \ref{eqGJES}).

The frequency resolution of a JES ($\delta\nu_{JES}$) can be calculated from \cite{Virtanen}\\
\begin{equation}
\delta\nu_{JES} = \dfrac{4}{\hbar}\sqrt{2 \ln{2} k_B {T_e}^2 {{C_{e,JES}}}}\text{.}
\label{FreqResSES}
\end{equation}
The electron heat capacitance needs to be calculated at the current-dependent escape temperature [$T_e(I)$], thus in the superconducting state, and takes the form\\
\begin{equation}
{C_{e,JES}} = \gamma_{A} \mathcal{V}_{A}T_e \Theta_{Damp}={C_{e,nano\text{-}TES}}\Theta_{Damp}\text{,}
\end{equation}
where the electronic heat capacitance is given by
\begin{equation}
C_s = 1.34 \gamma_{A} T_C^i \left( \dfrac{\Delta_A}{k_B T_e} \right)^{-3/2} e^{-\Delta_A/k_B T_e}\text{.}
\end{equation}
Furthermore, $\Theta_{Damp}$ is the low temperature exponential suppression with respect to the non-superconducting metal value, an it takes the form \cite{Rabani}\\
\begin{equation}
\Theta_{Damp} = \dfrac{C_s}{1.34\gamma_{A} T_e}\text{.}
\end{equation}

Since the JES does not operate at the middle of the superconducting-to-normal-state transition, the JES time response does not depend on the electrothermal parameter. Indeed, it is given by the relaxation half-time ($\tau_{1/2}$), which reads \cite{Virtanen}\\
\begin{equation}
\tau_{1/2} = \tau_{JES} \ln{2} \text{,}
\label{TauMezzi} 
\end{equation}
where $\tau_{JES}$ is the JES intrinsic thermal time constant. The latter is calculated by considering ${C_{e,JES}}$ and ${G_{th,JES}}$ in deep superconducting operation. Indeed, the JES parameters obtained in the experiments are in Eq. \ref{IntrinsicTime} to compute the response time in the superconducting state

\begin{table*}
\centering
\renewcommand\tabcolsep{11pt}
\begin{tabular}{c c c c c c c |c}
\noalign{\smallskip}\hline\noalign{\smallskip}

Sample & $\mathrm{T_c}$ & $\tau$ & $\tau_{eff}$ & $\delta\nu$ &   $\nu/\delta\nu$ \\ 
& (mK) & ($\mu s$) & ($\mu s$) & (GHz) &  100 GHz \quad 300 GHz \\
\noalign{\smallskip}\hline\noalign{\smallskip}
1 & 128 & 6 & 0.01 & 100 & 1 \qquad\qquad 3 \\
2 & 139 & 5 & 0.2 & 540 & 0.18 \qquad\quad 0.55\\
\noalign{\smallskip}\hline
\end{tabular}
\caption{\textbf{Main figures of merit deduced for the nano-TES.} The time constant $\tau$, the pulse recovery time $\tau_{eff}$, the frequency resolution $\delta \nu$,  and the resolving power $\nu/\delta\nu$ (at 100 and 300 GHz) are reported for the two fabricated nano-TESs.}
\label{tab:FigureOfMerit}
\end{table*}

\subsection{Performance deduced from the experimental data}
In this section, we show the sensing performance obtained for two different 1DJs (samples 1 and 2 of Fig. \ref{FigImod}) operated both as nano-TES and JES. 

Table \ref{tab:FigureOfMerit} summarizes the figures of merit calculated in case of nano-TES operation. The intrinsic relaxation time of the active region is limited by $G_{th,nano-TES}$ to a few microseconds for both devices ($\tau_1\simeq6\;\mu$s and $\tau_2\simeq5\;\mu$s). On the one hand, the electron heat capacitance is ${C_{e,nano-TES}}_1 =4 \times 10^{-20}$ J/K and the thermal conductance takes value ${G_{th,nano-TES}}_1= 6.7\times10^{-15}$ W/K for sample 1. On the other hand, ${C_{e,nano-TES}}_2 =4.2 \times 10^{-20}$ J/K and ${G_{th,nano-TES}}_2=9.3\times 10^{-15}$ W/K for sample 2. The thermal response of the nano-TES full detector is strongly damped by the eletrothermal parameter (since $\alpha\gg1$). In particular, the detector response time is $\tau_{eff,1} = 0.01\;\mu$s and $\tau_{eff,2} = 0.2\;\mu$s for sample 1 and sample 2, respectively. Indeed, the nano-TES response time is shorter than the intrinsic thermal response time of the superconducting thin film by more than one order of magnitude for both devices ($\tau_{eff} << \tau$). The frequency resolution depends on the electrothermal parameter ($\alpha^{-1/2}$), too. Therefore, the two nano-TESs show different values of $\delta\nu$. In particular, $\delta\nu_1\simeq$ 100 GHz ($\delta E_1\simeq$ 0.4 meV) for sample 1 and $\delta\nu_2\simeq$ 540 GHz ($\delta E_2 \simeq$ 2 meV) for sample 2 were calculated. Accordingly, the resolving power ($\nu/\delta \nu$) is larger than $1$ for $\nu \geq100$ GHz for sample 1. 

\begin{figure}[t!]
\centering
\includegraphics[width=\columnwidth]{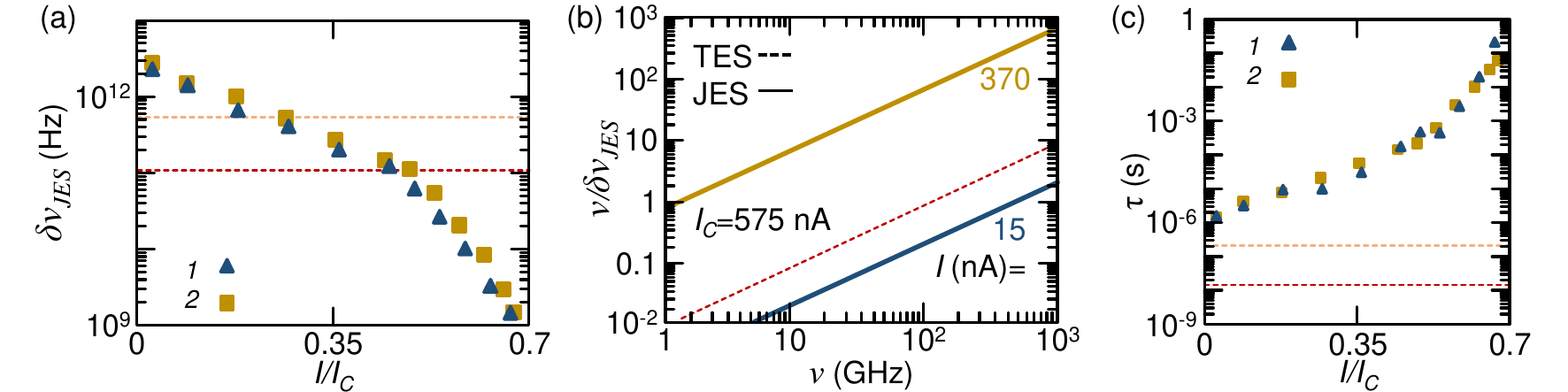}
\caption{\textbf{Performance calculated for the JES.} (\textbf{a}) Frequency resolution $\delta \nu_{JES}$ as a function of $I$ for sample $1$ (blue triangles) and $2$ (yellow squares). The dashed lines indicate the frequency resolutions of the corresponding nano-TESs.
(\textbf{b}) Resolving power ($\nu/ \delta \nu_{JES}$) as a function of the  frequency ($\nu$) of the incident single-photons calculated for sample $1$. The dashed lines indicate the resolving power of the corresponding nano-TES.
(\textbf{c}) Time constant versus bias current for sample 1 (blue triangles) and $2$ (yellow squares). The dashed lines indicate the time constants of the corresponding nano-TESs.}
\label{fig:Fig4}
\end{figure}

The performance in the JES operation are expected to strongly depend on the bias current. Indeed, Fig. \ref{fig:Fig4}(a) emphasizes the variations over $3$ orders of magnitude of $\delta \nu_{JES}$ on $I$. 
The best frequency resolution is $\sim 2$ GHz at $370$ nA. This value would enable the detection of single-photons at unprecedented low energies. The disrupting sensitivity is highlighted by the resolving power ($\nu/ \delta \nu_{JES}$). Figure \ref{fig:Fig4}(b) shows $\nu/ \delta \nu_{JES}$ calculated as a function of the frequency of the incident photons. In particular,
$\nu/ \delta \nu_{JES}$ can reach $\sim 80$ at $100$ GHz and  $ \sim 240$ at $300$ GHz for $370$ nA. 

The dependence of the JES time constant ($\tau_{1/2}$) on $I$ is shown in Fig. \ref{fig:Fig4}(c). In particular, $\tau_{1/2}$ monotonically increases by rising $I$, and varies between $\sim 1\,\mu$s at low current amplitude and $\sim 100\,$ms at $370$ nA. Notably, these values are orders of magnitude larger than the ones of nano-TESs. As a consequence, the read-out of single photons with the JES allows to employ slower and thus cheaper electronics.

Concluding, both the nano-TES and the JES show frequency resolutions enabling the search of axions through LSW experiments in the microwave frequency band. In particular, the JES allows to perform experiments in a wide range of energies down to about 8 $\mu$eV (2 GHz). Furthermore, the slow response time of the JES is not detrimental for LSW experiments. Indeed, approximately ten photon-axion-photon conversions are expected in one year \cite{Capparelli}, and a slower electronics can be used to read-out the detector response.

\section{Materials and Methods}
\label{Mater}
\subsection{Fabrication procedure}
All the devices presented in this review were fabricated by electron-beam lithography (EBL) and 3-angles shadow evaporation through a suspended resist mask onto a silicon wafer covered with 300-nm-thick SiO$_{2}$ thermally grown on an intrinsic silicon wafer. 
To obtain the resist suspended mask, a bilayer composed of a 950-nm-thick MMA(8.5)MMA layer and a PMMA (A4, 950k) film of thickness of about 300 nm was spin-coated on the substrate. The ratio between the electron irradiation doses to make the resists soluble is $DOSE_{MMA}:DOSE_{PMMA}\simeq1:4$. 
The evaporations were performed in an ultra-high vacuum electron-beam evaporator with a base pressure of about $10^{-11}$ Torr by keeping the target substrate at room temperature. First, 13-nm-thick Al layer was evaporated at an angle of -40$^\circ$. Second, the film was then oxidized by exposition to 200 mTorr of O$_{2}$ for 5 minutes to obtain the tunnel probes of the device devoted to the spectral and the thermal measurements. Third, the Al/Cu bilayer ($t_{Al} = 10.5$ nm and $t_{Cu} = 15$ nm) forming the superconducting nanowire is evaporated at an angle of $0^\circ$. Fourth, a second 40-nm-thick Al film was evaporated at an angle of $+40^\circ$ to obtain the lateral electrodes completing the 1DJ. The angle resolution of each evaporation was  $\sim1^\circ$.
The average film thickness can be controlled during the evaporation process with the precision of 0.1 nm at the evaporation rate of about 1.5 angstrom/s.

\subsection{Measurement setups}
The electronic and the spectral characterizations  presented in this review were performed at cryogenic temperatures in a $^3$He-$^4$He dilution refrigerator equipped with RC low-pass filters (cut-off frequency of about 800 Hz). The lowest electronic temperature obtained was 20 mK. 

The bias current tuning of the transport properties of the 1DJ is realized by standard lock-in technique. The AC current bias is produced by applying a voltage $V_{ac}$ at a frequency 13.33 Hz to a load resistance $R_{L}=100$ k$\Omega$ ($R_{l}>>R_N$) in order to obtain a bias current independent from the resistance of the 1DJ. The voltage drop $V$ across the device is measured as a function of $T_{bath}$ by means of a voltage pre-amplifier connected to a lock-in amplifier. The use of a pre-amplifier usually improves the signal to noise ratio. The control of the transport properties of the 1DJ is thus performed by varying $V_{ac}$. 

The energy gap of the superconducting nanowire was determined by tunnel spectroscopy. The voltage bias was applied between one tunnel probe and one lateral electrode by means of a low noise DC source, while the flowing current was measured through a room temperature current pre-amplifier.

\subsection{Basic properties of the active region}
The diffusion constant of the active region ($D_{A}$) is the average of the constants of the two components weighted on their thickness:  $D_{A}=(t_{Al}D_{Al}+t_{Cu}D_{Cu})/(t_{Al}+t_{Cu})\simeq5.6\times10^{-3}\text{m}^2/\text{s}$ (where $D_{Al}=2.25\times 10^{-3}$ m$^2$s$^{-1}$ for the Al thin film, and $D_{Cu}=8\times 10^{-3}$ m$^2$s$^{-1}$ for the Cu layer).
The superconducting coherence length of the Al film is $\xi_{Al}=\sqrt{\hbar D_{Al}/\Delta_{Al}}\simeq80$ nm, where $\Delta_{Al}\simeq200\;\mu$eV is its measured superconducting energy gap. The normal-state coherence length of the Cu layer is $\xi_{Cu}=\sqrt{\hbar D_{Cu}/(2\pi k_B T)} \simeq255$ nm, where $T=150$ mK is chosen in the worst possible working scenario (that is the critical temperature of $A$).
The superconducting coherence length in $A$ is given by $\xi_A=\sqrt{l \hbar/[(t_{Al}N_{Al}+t_{Cu}N_{Cu}) R_Ne^2\Delta_{A,0}]}\simeq 220$ nm, where $R_N=80\;\Omega$ is the nanowire normal-state resistance, $N_{Al}=2.15\times 10^{47}$ J$^{-1}$m$^{-3}$ and $N_{Cu}=1.56\times 10^{47}$ J$^{-1}$m$^{-3}$ are the density of states at the Fermi level of Al and Cu, respectively. The London penetration depth for the magnetic field of $A$ takes the form $\lambda_{L,A}=\sqrt{(\hbar wt_{A} R_N)/(\pi \mu_0 l \Delta_{A,0})}\simeq 970$ nm, where $\mu_0$ is the magnetic permeability of vacuum.
The magnetic field generated at the wire surface in correspondence of the maximum bias current is $B_{I,max}=\mu_0 I_{max}/(2\pi t)\simeq 4.7$ $\mu$T, where $I_{max}=370$ nA and $\mu_0$ is the vacuum magnetic permeability.
\section{Conclusions}
\label{Concl}
This paper reviewed two innovative hypersensitive superconducting radiation sensors: the nanoscale transition edge sensor (nano-TES) and the Josephson escape sensor (JES). Both devices are based on a one-dimensional Josephson junction (1DJ). This structure allows to in situ fine tune their performance by simple current bias. Indeed, the critical temperature ($T_C$) and the escape temperature ($T_e$) of a 1DJ can be mastered by controlling the flowing supercurrent. The nano-TES and the JES have the potential to detect single-photons in the gigahertz band towards unexplored levels of sensitivity. In fact, the nano-TES shows a frequency resolution of about 100 GHz, while the JES is able to resolve single-photons down to 2  GHz. Therefore, these sensors are the ideal candidate for the implementation of light shining through walls (LSW) experiments for the search of axions operating at micro- and milli-electronVolt energies. Furthermore, the nano-TES and the JES could have countless applications in several fields of quantum technology where single-photon detection is a fundamental task, such as quantum computation \cite{Brien} and quantum cryptography \cite{Gisin,Tittel}.

\funding{This research was funded by the European Union's Horizon 2020 research and innovation programme under the grant No. 777222 ATTRACT (Project T-CONVERSE) and under FET-Open grant agreement No. 800923-SUPERTED, and by the CSN V of INFN under the technology innovation grant SIMP.}

\acknowledgments{We acknowledge N. Ligato, G. Germanese, V. Buccheri, P. Virtanen, P. Spagnolo, C. Gatti, R. Paoletti, F.S. Bergeret, G. De Simoni, E. Strambini, A. Tartari, G. Signorelli and G. Lamanna for fruitful discussions.}

\conflictsofinterest{The authors declare no conflict of interest.} 



\reftitle{References}



\begin{thebibliography}{999}
\bibitem{Preskill}
Preskill, J., Wise, M. B., and Wilczek, F.,
Cosmology of the invisible axion, 
\emph{Phys. Lett. B} {\bf{1983}}, 120, 127-132.

\bibitem{Abbott}
Abbott, L. F., and Sikivie, P.,
A cosmological bound on the invisible axion, 
\emph{Phys. Lett. B} {\bf{1983}}, 120, 133-136.

\bibitem{Peccei}
Peccei, R. D., and Quinn, H., 
CP Conservation in the Presence of Instantons, 
\emph{Phys. Rev. Lett.} {\bf{1977}}, 38, 1440.

\bibitem{Dine}
Dine, M., Fischler, W., and Srednicki, M.,
A Simple Solution to the Strong CP Problem with a Harmless Axion, 
\emph{Phys. Lett. B} {\bf{1981}}, 104, 199.

\bibitem{Kim}
Kim, J. E., 
Weak Interaction Single and Strong CP Invariance, 
\emph{Phys. Rev. Lett.} {\bf{1979}}, 43, 103.

\bibitem{Zioutas}
Zioutas, K., Tsagri, M., Semertzidis, Y., Papaevangelou, T., Dafni, T., and Anastassopoulos, V. 
Axion searches with helioscopes and astrophysical signatures for axion(-like) particles,
\emph{New J. Phys.} {\bf{2009}}, 11, 105020.

\bibitem{Sikivie}
Sikivie, P., 
Experimental Tests of the "Invisible" Axion,
\emph{Phys. Rev. Lett.} {\bf{1983}}, 51, 1415.

\bibitem{Bibber}
van Bibber, K., McIntyre, P. M., Morris, D. E., and Raffelt, G. G.,
Design for a practical laboratory detector for solar axions,
\emph{Phys. Rev. D} {\bf{1989}}, 39, 2089.

\bibitem{Arik}
Arik, M., et al.,
New solar axion search using the CERN Axion Solar Telescope with 4He filling,
\emph{Phys. Rev. D} {\bf{2015}}, 92, 021101.

\bibitem{Armengaud}
Armengaud, E., et al.,
Conceptual Design of the International Axion Observatory (IAXO),
\emph{JINST} {\bf{2014}}, 9, T05002.

\bibitem{Asztalos}
Asztalos, S. J., et al., 
A SQUID-based microwave cavity search for dark-matter axions,
\emph{Phys. Rev. Lett.} {\bf{2010}}, 104, 041301.

\bibitem{Barbieri}
Barbieri, R., et al., 
Searching for galactic axions through magnetized media: The QUAX proposal,
\emph{Phys. Dark Universe} {\bf{2017}}, 15, 135-141.

\bibitem{Graham}
Graham, P. W., et al., 
Experimental Searches for the Axion and Axion-Like Particles,
\emph{Ann. Rev. Nuc. Part. Sci. } {\bf{2016}}, 65, 485-514.

\bibitem{Capparelli}
Capparelli, L. M., et al., 
Axion-like particle searches with sub-THz photons,
\emph{Phys. Dark Universe} {\bf{2016}}, 12, 37-44.

\bibitem{Alesini1}
Alesini, D., et al., 
Status of the SIMP Project: Toward the Single Microwave Photon Detection,
\emph{J Low Temp Phys} {\bf{2020}}, 199, 348–354.

\bibitem{Alesini2}
Alesini, D., et al., 
Development of a Josephson junction based single photon microwave detector for axion detection experiments,
\emph{J. Phys.: Conf. Ser.} {\bf{2020}}, 1559, 012020.

\bibitem{Irwin1995}
Irwin, K. D.,
An application of electrothermal feedback for high resolution cryogenic particle detection,
\emph{Appl. Phys. Lett.} {\bf{1995}}, 66, 1998-2000.

\bibitem{Irwin2006}
Irwin, K. D., 
Seeing with Superconductors, 
\emph{Sci. Am.} {\bf{2006}}, 295, 86-94.

\bibitem{Khosropanah}
Khosropanah, P., et \textit{al.},
Low noise transition edge sensor (TES) for the SAFARI Instrument on SPICA. 
\emph{Proc. SPIE 7741, Millimeter, Submillimeter, and Far-Infrared
Detectors and Instrumentation for Astronomy V} {\bf{2010}}, 77410L.

\bibitem{Visser}
de Visser, P., et \textit{al.},
Fluctuations in the electron system of a superconductor exposed to a photon flux, \emph{Nat Commun.} {\bf{2014}}, 5, 3130.

\bibitem{Monfardini} 
Monfardini, A., et \textit{al.},
Lumped element kinetic inductance detectors for space applications, 
\emph{Proc. SPIE 9914, Millimeter, Submillimeter, and Far-Infrared Detectors and Instrumentation for Astronomy VIII} {\bf{2016}}, 99140N.

\bibitem{Josephson}
Josephson, B. D.,
Possible new effects in superconductive tunnelling,
\emph{Phys. Lett.} {\bf{1962}}, 1, 251-253.

\bibitem{Solinas}
Solinas, P., Giazotto, F., and Pepe, G. P.,
Proximity SQUID Single-Photon Detector via Temperature-to-Voltage Conversion, 
\emph{Phys. Rev. Applied} {\bf{2018}}, 10, 024015.

\bibitem{Guarcello}
Guarcello, C., Braggio, A., Solinas, P., Pepe, G. P., and Giazotto, F., 
Josephson-Threshold Calorimeter, 
\emph{Phys. Rev. Applied} {\bf{2019}}, 11, 054074.

\bibitem{Giazotto2008}
Giazotto, F., Heikkil\"a, Pepe, G. P., Helist\"o, P., T. T., Luukanen, and Pekola, J. P.,
Ultrasensitive proximity Josephson sensor with kinetic inductance readout, 
\emph{Appl. Phys. Lett.} {\bf{2008}}, 92, 162507.

\bibitem{Virtanen}
Virtanen, P., Ronzani, A., and Giazotto, F.,
Josephson Photodetectors via Temperature-to-Phase Conversion, 
\emph{Phys. Rev. Applied} {\bf{2018}}, 9, 054027.

\bibitem{Natarajan}
Natarajan, C. M., Tanner, M. G., and Hdfield, R. H.,
Superconducting nanowire single-photon detectors: physics and applications, 
\emph{Supercond. Sci. Technol.} {\bf{2012}}, 25, 063001.

\bibitem{Oelsner}
Oelsner, G., et al., 
Underdamped Josephson junction as a switching current detector, 
\emph{Appl. Phys. Lett.} {\bf{2013}}, 103, 142605.

\bibitem{Oelsner2}
Oelsner, G., et al., 
Detection of weak microwave fields with an underdamped Josephson junction, 
\emph{Phys. Rev. Appl.} {\bf{2017}}, 7, 014012.

\bibitem{Kuzmin}
Kuzmin, L. S., et al., 
Single Photon Counter Based on a Josephson Junction at 14 GHz for Searching Galactic Axions, 
\emph{IEEE Trans. Appl. Supercond.} {\bf{2018}}, 28, 2400505.

\bibitem{Paolucci2}
Paolucci, F., Buccheri, V., Germanese, G., Ligato, N., Paoletti, R., Signorelli, G., Bitossi., M., Spagnolo, P., Falferi, P., Rajteri, M., Gatti, C., and Giazotto, F.,
Highly sensitive nano-TESs for gigahertz astronomy and dark matter search
\emph{J. Appl. Phys.}, {\bf{2020}}, 128, 194502.

\bibitem{Paolucci}
Paolucci, F., Ligato, N., Buccheri, V., Germanese, G., Virtanen, P., and Giazotto, F.,
Hypersensitive tunable Josephson escape sensor for gigahertz astronomy, 
\emph{Phys. Rev. Applied}, {\bf{2020}}, 14, 034055.

\bibitem{tinkham}
Tinkham, M., \emph{Introduction to Superconductivity}, McGraw-Hill: New York, USA, 1996.

\bibitem{Barone1982}
Barone, A., and Patern\`o, G.,
\emph{Physics and Applications of the Josephson Effect}, Wiley-VCH: New York, USA, 1982.

\bibitem{Bezryadin2012}
Bezryadin, A.,
\textit{Superconductivity in Nanowires: Fabrication and Quantum Transport} Wiley-VCH: New York, USA, 2012.  

\bibitem{Ivanchenko1969}
Ivanchenko, Yu. M., and Zil'berman, L. A.,
The Josephson effect in small tunnel contacts,
\emph{Sov. Phys. JETP} {\bf{1969}}, 28, 1272.

\bibitem{cochran}
Cochran, J. F.,  and Mapother, D. E.,
Superconducting Transition in Aluminum, 
\emph{Phys Rev.} {\bf{1958}}, 111, 132.

\bibitem{Giazotto2006}
Giazotto, F., Heikkil\"a, T. T., Luukanen, A., Savin, A. M., and Pekola, J. P.,
Opportunities for mesoscopics in thermometry and refrigeration: Physics and applications,
\emph{Rev. Mod. Phys.} {\bf{2006}}, 78, 217-274.

\bibitem{Golubov}
Golubov, A. A., Kupriyanov, M. Yu., and Il’ichev, E.,
The current-phase relation in Josephson junctions,
\emph{Rev. Mod. Phys.} {\bf{2004}}, 76, 411-479.

\bibitem{DeGennes1964}
De Gennes, P. G.,
Boundary Effects in Superconductors,
\emph{Rev. Mod. Phys.} {\bf{1964}}, 36, 225.

\bibitem{Kogan1982}
Kogan, V. G.,
Coherence length of a normal metal in a proximity system,
\emph{Phys. Rev. B} {\bf{1982}}, 26, 88.

\bibitem{Courtois}
Courtois, H., Meschke, M., Peltonen, J. T., and Pekola, J. P.,
Hysteresis in a Proximity Josephson Junction,
\emph{Phys. Rev. Lett.} {\bf{2008}}, 101, 067002.

\bibitem{Andreev1964}
Andreev, A. F.,
The Thermal Conductivity of the Intermediate State in Superconductors,
\emph{JETP} {\bf{1964}}, 66, 1228-1231.

\bibitem{Timofeev2009}
Timofeev, A. V., Pascual Giarc\'ia, C., Kopnin, N. B., Savin, A. M., Meschke, M., Giazotto, F., and Pekola, J. P.,
Recombination-Limited Energy Relaxation in a Bardeen-Cooper-Schrieffer Superconductor,
\emph{Phys. Rev. Lett.} {\bf{2009}}, 102, 017003.

\bibitem{Heikkila}
Heikkil$\ddot{a}$, T. T., Silaev, M., Virtanen, P., and Bergeret, F. S.,
Thermal, electric and spin transport in superconductor/ferromagnetic-insulator structures, 
\emph{Progress in Surface Science} {\bf{2019}}, 94, 100540.

\bibitem{Rabani}
Rabani, H., Taddei, F., Bourgeois, O., Fazio, R., and Giazotto, F.,
Phase-dependent electronic specific heat in mesoscopic josephson junction,
\emph{Phys. Rev. B} {\bf{2008}}, 78, 012503. 

\bibitem{Brien}
O’brien, J. L.,
Optical Quantum Computing,
\emph{Science} {\bf{2007}}, 318, 1567-1570. 

\bibitem{Gisin}
Gisin, N., Ribordy, G., Tittel, W., and Zbiden, H.,
Quantum cryptography,
\emph{Rev. Mod. Phys.} {\bf{2002}}, 74, 145-195. 

\bibitem{Tittel}
Tittel, W.,
Quantum key distribution breaking limits,
\emph{Nat. Photonics} {\bf{2019}}, 13, 310-311. 

\end{thebibliography}



\end{document}